\documentclass[11pt]{article}

\usepackage{rotating}
\usepackage{sectsty}
\usepackage{longtable}
\usepackage{cite}
\usepackage{hyperref}
\usepackage{authblk}

\usepackage{graphicx}%
\usepackage{multirow}%
\usepackage{mathrsfs}%
\usepackage[title]{appendix}%
\usepackage{xcolor}%
\usepackage{textcomp}%
\usepackage{manyfoot}%
\usepackage{booktabs}%
\usepackage{algpseudocode}%
\usepackage{algorithm}%
\usepackage{algorithmicx}%
\usepackage{listings}%
\usepackage{makecell}
\usepackage{array}
\usepackage{booktabs}
\usepackage{bbding}
\usepackage{xcolor, colortbl}

\usepackage[singlelinecheck=off]{caption}
\usepackage{rotating}
\usepackage{float}
\usepackage{tabto}
\usepackage{tabularx}
\usepackage{url}
\usepackage{amstext}

\providecommand{\keywords}[1]
{
	\small	
	\textbf{Keywords \space} #1
}

\title{Multilevel User Credibility Assessment in Social Networks}

\author[1]{Mohammad Moradi}
\author[1]{Mostafa Haghir Chehreghani}
\affil[1]{Department of Computer Engineering, Amirkabir University of Technology (Tehran Polytechnic), Tehran, Iran}

\date{}

\begin{document}
	\maketitle
	\begin{abstract}
Online social networks are major platforms for disseminating both real and false information. Many users--intentionally or unintentionally--spread harmful content, misinformation, and rumors in domains such as politics and business. Consequently, user credibility assessment has become an important research area.  
Most existing methods suffer from two limitations. First, they treat credibility as a binary label (genuine or fake), while real-world applications often require a more nuanced, multilevel evaluation. Second, they use only a subset of relevant features, which limits predictive performance.  
To address these shortcomings, we first create a dataset tailored for multilevel credibility assessment. We then propose MultiCred, a model that assigns users to multiple credibility tiers using a rich set of features extracted from profiles, tweets, and comments. MultiCred combines deep language models for textual analysis with neural networks for non-textual feature processing.  
Extensive experiments show that MultiCred significantly outperforms existing approaches across multiple evaluation metrics.
	\end{abstract}
	
	
	
\section{Introduction }
\label{Introduction}


Social networks have become indispensable platforms for information exchange, opinion formation, and news dissemination due to their accessibility, low cost, and vast user base. However, their openness also facilitates the rapid spread of misinformation, rumors, and malicious content. False information shared intentionally or inadvertently by untrustworthy accounts can distort public perception, influence elections, and undermine crisis management efforts. The ability to reliably assess the credibility of users on such platforms is therefore crucial for maintaining information integrity, fostering healthy online discourse, and supporting downstream applications such as content moderation, recommendation systems, and trust-aware information retrieval.

Previous research on user credibility and fake-account detection in social networks has explored diverse perspectives and methodologies. Early work concentrated on profile-based features and handcrafted heuristics (e.g., follower-following ratios, posting behavior, profile-image checks) to separate fake from genuine accounts \cite{singh2018detection,DBLP:conf/comsnets/WaniAJH19,DBLP:conf/iscc/ZareiFC19}. Parallel efforts focused on textual signals, leveraging linguistic, sentiment and topic-level features to detect automation, spam, and malicious content \cite{DBLP:conf/ACISicis/SweM18,DBLP:journals/jocs/ClarkWJGDD16,DBLP:journals/tdsc/KhanAAKZ18,DBLP:conf/icccnt/PhadC18}. More recent approaches fuse textual, profile and network information and apply machine learning \textendash deep learning classifiers (including hybrid ML-DL pipelines) to attain higher detection accuracy \cite{ala2017spam,DBLP:conf/asunam/AlomCF18,DBLP:journals/isf/AswaniKI18,DBLP:journals/mta/AdewoleAKS19,DBLP:conf/bigdataconf/AsghariCC22,DBLP:journals/snam/VermaAMG22}. Several studies have also explored specialized signals such as emotion-sentiment patterns \cite{DBLP:conf/comsnets/WaniAJH19}, impersonation and political actor impersonators \cite{DBLP:conf/iscc/ZareiFC19}, and deep-profile image-text fusion models \cite{DBLP:journals/istr/WandaJ20}, which further highlight the diversity of useful indicators. Despite these advances, most methods still treat credibility as a binary problem (fake vs. genuine) and often rely on a limited subset of features or single-domain signals\textemdash an oversimplification that neglects the continuum of credibility behaviors found in real-world networks.

As a result,
existing approaches remain limited in two crucial aspects. First, they predominantly model user credibility as a binary problem, overlooking the nuanced spectrum of user behaviors that range from highly credible to intentionally deceptive. This simplification fails to capture intermediate cases, such as users who occasionally share misleading content or amplify unverified information without malicious intent. Second, there is no publicly available dataset that supports a fine-grained, multilevel characterization of user credibility, which hinders progress toward more realistic modeling and benchmarking. Addressing these limitations requires both an appropriately labeled dataset and a framework capable of integrating diverse feature types to assess credibility at multiple levels.


To overcome these limitations, we introduce MultiCred, a comprehensive framework for multilevel user credibility assessment on social networks. Our approach begins by constructing a labeled dataset that categorizes Twitter users across several credibility tiers, capturing the full behavioral spectrum from highly reliable to deceptive accounts. The MultiCred model then integrates three complementary sources of information: profile features that reflect user identity and social activity, content features that represent linguistic and semantic characteristics of posts, and interaction features that encode how other users perceive and engage with the account. By employing specialized preprocessing and analysis pipelines for each feature group and combining their outputs within a unified learning architecture, MultiCred can capture diverse credibility signals and generalize across different user types. This design enables a more nuanced and scalable understanding of user trustworthiness compared to existing binary classification methods.

We evaluate MultiCred on a collected real-world Twitter dataset and compare it against two state-of-the-art credibility-assessment models, UCred \cite{DBLP:journals/snam/VermaAMG22} and the model of \cite{DBLP:journals/soco/BhartiP21}. In the scenario where UCred performs best, MultiCred achieves a 
15.07\% improvement in F1 score, demonstrating a substantial advantage in distinguishing users across credibility levels. Additionally, MultiCred considerably outperforms the model of \cite{DBLP:journals/soco/BhartiP21}. These results confirm the effectiveness of integrating profile, content, and interaction features for robust multilevel user credibility assessment.

The rest of this paper is organized as follows. Section~\ref{Related work} reviews prior research in the area. Section~\ref{Preliminaries} introduces fundamental concepts and formalizes the problem. Section~\ref{Dataset} describes our data-collection procedure and the resulting dataset. Section~\ref{Proposed method} details the proposed algorithm and its components. Section~\ref{Empirical results} presents the experimental findings. Finally, Section~\ref{Conclusion and future work} concludes the paper and outlines directions for future research.

\section{Related work} \label{Related work}

Over the past decade, social networks have attracted global attention, leading platforms such as Facebook, Twitter, LinkedIn, and Instagram to experience a dramatic surge in user registrations. However, not all accounts are genuine; many are created for malicious or strategic purposes. Recent studies have employed a range of advanced techniques to identify these fake accounts. Broadly speaking, existing methods for fake-account detection can be divided into three main categories:

\begin{itemize}
	\item Methods using non-textual (profile-based) features.
	\item Methods using textual features.
	\item Methods combining both textual and non-textual features.
\end{itemize}

In the remainder of this section, we review representative approaches from each category.

\subsection{Methods based on user profile features}

Singh et al. \cite{singh2018detection} employed supervised machine learning models to detect fake profiles on social networks. They used the average follower count to distinguish between fake and genuine accounts, finding that profiles with more than 30 followers are unlikely to be fake. They also discovered that the average age of fake-profile owners is between 18 and 19 years, and that their profile images are often sourced from the internet.

Agarwal et al. \cite{DBLP:conf/comsnets/WaniAJH19} proposed a supervised model for detecting fake Facebook accounts by analyzing user sentiment.
They extracted emotion-based features--including anger, sadness, fear, joy, trust, and the frequencies of positive and negative sentiment terms. Their analysis revealed that users behind fake profiles predominantly express emotions related to hatred, violence, and ugliness.

Zarei et al. \cite{DBLP:conf/iscc/ZareiFC19} proposed a model for detecting fake political accounts on social media. They collected data from three politicians' Instagram profiles and used it to identify a substantial number of fake individuals and political bots. The authors noted that this was the first study to perform such an analysis on Instagram data. They applied the TF-IDF technique to detect accounts with similar profile information and employed convolutional neural networks to compare profile images.

Wanda and Jie \cite{DBLP:journals/istr/WandaJ20} introduced a deep neural model called DeepProfile for detecting fake accounts on online social networks. They improved classification accuracy by modifying the pooling layer in the convolutional neural network.
Kumari et al. \cite{DBLP:journals/soco/BhartiP21} designed a system for identifying fake users on Twitter. Since we use their method as one of our baselines, we discuss it in detail in Section~\ref{Empirical results}.

\subsection{Methods based on textual features}

Swe and Myo \cite{DBLP:conf/ACISicis/SweM18} introduced a blacklist-based method for detecting fake accounts on online social networks. They generate the blacklist using topic modeling and keyword extraction, eliminating the need for profile- or network-based features and thus reducing the time and cost of feature extraction. The authors evaluated their approach on the 1KS-10KN and Honeypot datasets.

Clark et al. \cite{DBLP:journals/jocs/ClarkWJGDD16} applied natural language processing to automate bot detection on Twitter. Their model uses human-generated text to establish criteria for identifying accounts that post automated messages. They collected two datasets: geolocated tweets from 1,000 active users (the Geo-Tweet dataset) for human-versus-bot classification, and a dataset of honeypot content. Their findings indicate that model accuracy on the Geo-Tweet dataset increases with tweet length.

Khan et al. \cite{DBLP:journals/tdsc/KhanAAKZ18} distinguished spammers and bloggers from genuine domain experts on Twitter. They collected approximately 0.4 million tweets from about 3,200 users who actively share health-related information. To identify spammers and bloggers, they employed the link-based HITS topic search algorithm, differentiating them from domain experts. Their approach requires minimal pre-labeled data to distinguish fake users from genuine experts.

Phad and Chavan \cite{DBLP:conf/icccnt/PhadC18} proposed a model for detecting compromised profiles on social networks. They retrieved data from the Twitter Archive, compiling 26,363 tweets from 48 prominent accounts, of which 1,000 were malicious. Their model constructs a behavioral history for each user and assesses whether an account is at risk based on deviations from its normal activity patterns.

\subsection{Methods based on both textual and non-textual features}

Al-Zoubi et al.\ \cite{ala2017spam} identified spam profiles on Twitter using general-purpose features. They compiled a dataset of 82 user profiles that post in English and Arabic. Extracted features included the presence of suspicious words, use of the default profile picture, text-to-link ratio, comment ratio, and tweeting times. They employed machine learning classifiers--decision tree, C4.5, k-nearest neighbors, Naive Bayes, and multi-layer perceptron--to classify profiles as spam or non-spam.

Alom et al.\ \cite{DBLP:conf/asunam/AlomCF18} proposed a model for detecting spam accounts on Twitter by combining graphical and content-based features. They evaluated several classifiers--including k-nearest neighbors, decision tree, Naive Bayes, random forest, logistic regression, support vector machine (SVM), and XGBoost--on these features to distinguish spam accounts from legitimate ones.

Aswani et al. \cite{DBLP:journals/isf/AswaniKI18} proposed a model to identify spammers on Twitter. They collected 1,844,701 tweets from 14,235 user profiles and extracted 13 statistical features from social media analytics. They applied a bio-inspired Firefly algorithm to distinguish spammers from regular users.
Adewole et al. \cite{DBLP:journals/mta/AdewoleAKS19} developed a model for detecting both spam messages and spam accounts on online social networks. For spam message detection, they used three datasets: SMS Collection V.1 (5,574 samples), SMS Corpus V.0.1 Big (1,324 samples), and the Twitter Spam Corpus (18,000 samples). They extracted 18 features and evaluated various machine learning algorithms, with Random Forest achieving the best performance.

Asghari et al.~\cite{DBLP:conf/bigdataconf/AsghariCC22} explored various network measures--including centrality indices \cite{DBLP:conf/cikm/ChehreghaniBA19} and their correlations--to distinguish real users from fake ones. They showed that metrics such as average path length, eigenvector centrality, harmonic centrality, degree, and local reaching centrality, together with their pairwise correlations, serve as strong indicators for separating genuine and fake accounts.
Verma et al.~\cite{DBLP:journals/snam/VermaAMG22} proposed a method for evaluating Twitter user credibility using machine learning and deep learning techniques. As we adopt their approach as one of our baselines, we discuss it in detail in Section~\ref{Empirical results}.

Table~\ref{Overview} summarizes the reviewed works by classification type, features used, methodology, dataset source, and feature-engineering approach to highlight similarities and differences among prior studies.

\begin{table}[h]
	\caption{Overview of approaches to fake-user detection.\label{Overview}} 
	{ \tiny
		\begin{tabularx}{\textwidth}{c c c c c c@{\hspace{2pt}} c}
			\toprule
			\addlinespace[2pt]
			Method & \makecell{\textbf{Classification} \\ \textbf{(binary / multi-class)}} & \makecell{\textbf{Textual} \\ \textbf{features}} & \makecell{\textbf{Non-textual} \\ \textbf{features}} & \textbf{Methodology} & \makecell{\textbf{Dataset} \\ \textbf{source}} & \makecell{\textbf{Feature} \\ \textbf{engineering}} \\
			\addlinespace[2pt]
			\hline
			\addlinespace[2pt]
			\cite{singh2018detection} & binary & \XSolidBrush & \Checkmark & machine learning & collected & \XSolidBrush \\
			\addlinespace[2pt]
			\hline
			\addlinespace[2pt]
			\cite{DBLP:conf/comsnets/WaniAJH19} & binary & \XSolidBrush & \Checkmark & machine learning & collected & \Checkmark \\
			\addlinespace[2pt]
			\hline
			\addlinespace[2pt]
			\cite{DBLP:conf/iscc/ZareiFC19} & binary & \XSolidBrush & \Checkmark & \makecell{experimental \\ algorithm} & collected & \Checkmark \\
			\addlinespace[2pt]
			\hline
			\addlinespace[2pt]
			\cite{DBLP:journals/istr/WandaJ20} & binary & \XSolidBrush & \Checkmark & deep learning & collected & \XSolidBrush \\
			\addlinespace[2pt]
			\hline
			\addlinespace[2pt]
			\cite{DBLP:journals/soco/BhartiP21} & binary & \XSolidBrush & \Checkmark & \makecell{logistic regression \\ with PSO} & both & \Checkmark \\
			\addlinespace[2pt]
			\hline
			\addlinespace[2pt]
			\cite{DBLP:conf/ACISicis/SweM18} & binary & \Checkmark & \XSolidBrush & machine learning & pre-existing & \XSolidBrush \\
			\addlinespace[2pt]
			\hline
			\addlinespace[2pt]
			\cite{DBLP:journals/jocs/ClarkWJGDD16} & multi-class & \Checkmark & \XSolidBrush & \makecell{experimental \\ algorithm} & both & \XSolidBrush \\
			\addlinespace[2pt]
			\hline
			\addlinespace[2pt]
			\cite{DBLP:journals/tdsc/KhanAAKZ18} & binary & \Checkmark & \XSolidBrush & machine learning & collected & \Checkmark \\
			\addlinespace[2pt]
			\hline
			\addlinespace[2pt]
			\cite{DBLP:conf/icccnt/PhadC18} & binary & \Checkmark & \XSolidBrush & \makecell{experimental \\ algorithm} & collected & \XSolidBrush \\
			\addlinespace[2pt]
			\hline
			\addlinespace[2pt]
			\cite{ala2017spam} & binary & \Checkmark & \Checkmark & machine learning & collected & \Checkmark \\
			\addlinespace[2pt]
			\hline
			\addlinespace[2pt]
			\cite{DBLP:conf/asunam/AlomCF18} & binary & \Checkmark & \Checkmark & machine learning & pre-existing & \Checkmark \\
			\addlinespace[2pt]
			\hline
			\addlinespace[2pt]
			\cite{DBLP:journals/isf/AswaniKI18} & binary & \Checkmark & \Checkmark & machine learning & collected & \Checkmark \\
			\addlinespace[2pt]
			\hline
			\addlinespace[2pt]
			\cite{DBLP:journals/mta/AdewoleAKS19} & binary & \Checkmark & \Checkmark & machine learning & both & \Checkmark \\
			\addlinespace[2pt]
			\hline
			\addlinespace[2pt]
			\cite{DBLP:journals/snam/VermaAMG22} & binary & \Checkmark & \Checkmark & hybrid & pre-existing & \Checkmark \\
			\addlinespace[2pt]
			\hline
			\addlinespace[2pt]
			MultiCred (ours) & multi-class & \Checkmark & \Checkmark & deep learning & collected & \Checkmark \\
			\bottomrule
	\end{tabularx} }
\end{table}

As Table~\ref{Overview} shows, most prior studies treat fake-user detection as a binary classification task, with only one extending it to three classes. The majority rely primarily on textual features, while few integrate additional information sources. Moreover, only two studies employ deep learning; the remainder use traditional machine-learning approaches. Many works construct custom datasets--either created from scratch or by augmenting existing data--to evaluate their models. In contrast, our work combines textual and profile-based features within a deep-learning framework and moves beyond binary classification to assess multiple credibility levels. To this end, we collected a novel dataset annotated with varying degrees of user credibility, enabling a more fine-grained and realistic evaluation.

\section{Preliminaries} \label{Preliminaries}

In general, each user profile in a social network is described by a set of features--username, profile picture, description, and so on. We denote this set as 
\[
F = \{f_1, f_2, \ldots, f_n\},
\]
where \(f_i\) represents the \(i\)-th feature. These features span different modalities: the username is textual, the account creation time is numerical, and the profile picture is visual. Hence, it is crucial to integrate them appropriately at the outset. 

To this end, we define a mapping function \(z\) that transforms the feature set \(F\) into a fixed-dimensional vector while preserving the profile's essential information. Because feature types vary, \(z\) is tailored to each modality: numeric features use the identity mapping, whereas text and image features are encoded with models such as BERT \cite{DBLP:conf/naacl/DevlinCLT19} and CNNs \cite{fukushima:neocognitronbc}. 
Table~\ref{Notation} summarizes the main notation used in this section and throughout the paper.

\begin{table}[h]
	\caption{Notation.\label{Notation}} 
	{ \footnotesize
		\begin{tabularx}{\textwidth}{c c l }
			\toprule
			\addlinespace[2pt]
			Symbol & Section & Description \\
			\addlinespace[2pt]
			\hline
			\addlinespace[2pt]
			$F$ & \ref{Preliminaries} & set of features \\
			\addlinespace[2pt]
			$f_i$ & \ref{Preliminaries} & $i$-th feature \\
			\addlinespace[2pt]
			$z$ & \ref{Preliminaries} & mapping from feature space to vector space \\
			\addlinespace[2pt]
			$g$ & \ref{Preliminaries} & mapping from vector space to credibility label \\
			\addlinespace[2pt]
			$y_i$ & \ref{Preliminaries} & true class of the $i$-th data point \\
			\addlinespace[2pt]
			$p_i$ & \ref{Preliminaries} & predicted class probability vector for the $i$-th data point \\
			\addlinespace[2pt]
			$M$ & \ref{Preliminaries} & number of classes \\
			\addlinespace[2pt]
			$y_{i,c}$ & \ref{Preliminaries} & indicator whether the $i$-th data point belongs to class $c$ \\
			\addlinespace[2pt]
			$p_{i,c}$ & \ref{Preliminaries} & predicted probability that the $i$-th data point belongs to class $c$ \\
			\addlinespace[2pt]
			$\bar{x}$ & \ref{Non-textual features} & normalized value of data point $x$ \\
			\addlinespace[2pt]
			$k$ & \ref{Labels and class imbalance} & number of nearest neighbours \\
			\addlinespace[2pt]
			$\delta$ & \ref{Labels and class imbalance} & random variable in $(0,1)$ \\
			\addlinespace[2pt]
			$U$ & \ref{Labels and class imbalance} & uniform distribution \\
			\addlinespace[2pt]
			$d$ & \ref{Time Complexity} & number of numeric features \\
			\addlinespace[2pt]
			$m$ & \ref{Time Complexity} & number of tweets per user \\
			\addlinespace[2pt]
			$r$ & \ref{Time Complexity} & number of comments per user \\
			\addlinespace[2pt]
			$L_t$, $L_c$ & \ref{Time Complexity} & tweet length and comment length (in tokens) \\
			\addlinespace[2pt]
			$H$, $H_c$ & \ref{Time Complexity} & hidden size of BERT / DistilBERT representations \\
			\addlinespace[2pt]
			$C$ & \ref{Time Complexity} & compressed embedding size of the autoencoder \\
			\addlinespace[2pt]
			$P$ & \ref{Time Complexity} & total parameters in the classifier neural network \\
			\addlinespace[2pt]
			\bottomrule
	\end{tabularx} }
\end{table}

After mapping the feature set into vector space, we assess user credibility. Existing approaches typically treat credibility as a binary label--fake or genuine--which discards important gradations. Many ostensibly genuine users engage in harmful behavior on social networks, either knowingly or unknowingly, and such behavior should affect their credibility. Classifying users into only two categories therefore fails to capture these nuances. Defining multiple credibility levels yields a more precise and realistic assessment of each user.

Depending on the collected data, the number of credibility levels can be chosen. More levels yield a finer-grained understanding of credibility and its effects on user behavior. Once the number of levels is fixed, we define a function \(g\) that maps each feature vector to one of these levels. Our objective is to learn the mappings \(z\) and \(g\) so they accurately predict users' credibility.

To train these mappings, we employ an loss function--commonly the cross-entropy loss--which measures the divergence between the predicted probability distribution and the true distribution. The loss increases as the predicted probability deviates from the actual label. For binary classification, the cross-entropy loss for a single sample is defined as:
\begin{equation}
	\mathrm{CE} = -\bigl(y_i \log(p_i) + (1 - y_i) \log(1 - p_i)\bigr),
\end{equation}
where \(y_i\in\{0,1\}\) is the true class label of the \(i\)-th sample and \(p_i\in(0,1)\) is the predicted probability for that sample.

When handling multiclass classification, we use the categorical cross-entropy loss, which sums the individual class errors for each sample. It is defined as:
\begin{equation}
	\mathrm{CEE} = -\sum_{c=1}^{M} y_{i,c} \log\bigl(p_{i,c}\bigr),
\end{equation}
where \(M\) is the number of classes, \(y_{i,c}\in\{0,1\}\) indicates whether the \(i\)th sample belongs to class \(c\), and \(p_{i,c}\) is the predicted probability that the \(i\)th sample belongs to class \(c\). The overall loss for the model is the sum of these values over all training samples.

\section{Dataset} \label{Dataset}

This section first reviews existing datasets for user credibility assessment, then highlights their limitations, and finally describes our data-collection procedure and the resulting dataset.

\subsection{Existing datasets} \label{Available dataset}

Most datasets for fake-news and fake-user detection adopt a two-class scheme (fake vs. real), though some use three or five classes; all treat the problem as a classification task.
The InstaFake dataset \cite{akyon2019instagram} was developed to distinguish fake user accounts from automatically generated accounts (bots). For each class, a set of profile-based features was collected, including number of posts, number of followers, and the user biography. Textual content from posts and comments was not included in the dataset.

The FakeUserProfile dataset \cite{fameforsale2015} was collected from Twitter and contains data from 6,827 user accounts, of which 3,475 are labeled real and 3,352 labeled fake. All features were retrieved via the Twitter API, producing some attributes that are unique to this dataset. For example, properties such as \texttt{profile\_text\_color} and \texttt{profile\_sidebar\_border\_color}, along with other graphics-related attributes, are typically found only in Twitter-based datasets.

Several datasets have been developed for fake-news detection, including FakeNewsNet \cite{shu2019fakenewsnet} and LIAR-PLUS \cite{alhindi2018your}. Comprehensive reviews of fake-news datasets and detection algorithms are available in \cite{DBLP:journals/peerj-cs/DUliziaCFG21,DBLP:journals/air/LakzaeiCB24,DBLP:journals/corr/abs-2402-08401,lakzaei2025decisionbasedheterogenousgraphattention,DBLP:journals/corr/abs-2502-06927}.

\subsection{Limitations of existing datasets}
\label{Limitation with existing datasets}

The existing datasets have two primary limitations. First, each relies on a specific set of features for fake-user detection and omits other potentially informative attributes. Second, they support only binary classification and do not account for multiple credibility levels. Therefore, we collected a new dataset for this research rather than using the available ones.

\subsection{Our data collection method}
\label{Data collection}

We collect data from Twitter and label user accounts using NewsGuard \cite{newsguard}. NewsGuard evaluates news websites using scores from 0 to 100 based on multiple criteria; experienced reviewers and journalists perform these evaluations without the use of artificial intelligence. The frequency of spreading false information is the most important criterion. Table~\ref{NewsGuard} lists all criteria and their associated scores. The top five criteria primarily reflect the credibility of a news source, while the remaining four focus mainly on transparency and management. Therefore, the first five criteria are especially useful for assessing credibility, and the next four help evaluate transparency.

\begin{table}[h]
	\caption{NewsGuard evaluation criteria and scores.\label{NewsGuard}} 
	{ \footnotesize
		\begin{tabularx}{\textwidth}{l l c}
			\toprule
			\addlinespace[2pt]
			& Criterion & \hspace{0.7cm} Score \\
			\addlinespace[2pt]
			\hline
			\addlinespace[2pt]
			& Does not repeatedly publish false content. & \hspace{0.7cm} $22$ \\
			\addlinespace[2pt]
			& Gathers and presents information responsibly. & \hspace{0.7cm} $18$ \\
			\addlinespace[2pt]
			& Regularly corrects or clarifies errors. & \hspace{0.7cm} $12.5$ \\
			\addlinespace[2pt]
			& Handles the difference between news and opinion responsibly. & \hspace{0.7cm} $12.5$ \\
			\addlinespace[2pt]
			& Avoids deceptive headlines. & \hspace{0.7cm} $10$ \\
			\addlinespace[2pt]
			& Website discloses ownership and financing. & \hspace{0.7cm} $7.5$ \\
			\addlinespace[2pt]
			& Clearly labels advertising. & \hspace{0.7cm} $7.5$ \\
			\addlinespace[2pt]
			& Reveals who is in charge, including possible conflicts of interest. & \hspace{0.7cm} $5$ \\
			\addlinespace[2pt]
			& Provides the names of content creators and contact or biographical information. & \hspace{0.7cm} $5$ \\
			\addlinespace[2pt]
			\bottomrule
	\end{tabularx} }
\end{table}

Each user account on Twitter and other social networks typically shares posts on a variety of topics. For thematic categorization used to evaluate an account by content, posts about daily life and personal matters are of limited relevance and have little effect on the assessment. By contrast, posts that report news, describe events, or can be labeled as newsworthy are highly informative for determining a user's credibility. The aim of a credibility-assessment system is to let readers gauge the reliability of a user's activity at a glance. When a user shares a news item, the score should help readers make an informed judgment about that item's trustworthiness. For these reasons, the data-collection procedure is as follows:
\begin{itemize}
	\item In the first stage, we compile a list of news websites reviewed and scored by NewsGuard. These are primarily English-language sites based in the United States and Europe.
	\item In the second stage, we check each site for an official Twitter account and record its username. After the first two stages, the set of Twitter accounts to be queried is fully identified.
	\item In the third stage, we use the Python Tweepy library and the Twitter API to collect account data in three phases: first profile metadata, then user tweets, and finally user comments.
\end{itemize}

The primary focus of this research is user profile information. Consequently, in the initial phase we retrieve all profile attributes. Table~\ref{user profile feature} lists the profile-related features we extract. Twitter, like other social networks, awards a blue verification badge (the "blue tick") to certain accounts after review to indicate authenticity. This badge is typically granted to prominent individuals, public figures, and other influential accounts; the "verified" attribute therefore denotes profile authentication.

\begin{table}[h]
	\caption{Extracted features for each user profile.\label{user profile feature}}
	\begin{tabularx}{\textwidth}{l l l}
		\toprule
		\addlinespace[2pt]
		\multicolumn{3}{c}{Features} \\
		\addlinespace[2pt]
		\hline
		\addlinespace[2pt]
		name & screen\_name & description \\
		url & entities & profile\_image\_url \\
		profile\_image\_url\_https & profile\_banner\_url & profile\_background\_image\_url \\
		profile\_background\_image\_url\_https & profile\_background\_tile & profile\_use\_background\_image \\
		profile\_background\_color & profile\_text\_color & profile\_link\_color \\
		profile\_sidebar\_border\_color & profile\_sidebar\_fill\_color & default\_profile \\
		default\_profile\_image & protected & verified \\
		followers\_count & friends\_count & listed\_count \\
		favourites\_count & statuses\_count & created\_at \\
		utc\_offset & time\_zone & lang \\
		is\_translation\_enabled & translator\_type & contributors\_enabled \\
		geo\_enabled & location & profile\_location \\
		notifications & following & follow\_request\_sent \\
		withheld\_in\_countries & status & \\
		\addlinespace[2pt]
		\bottomrule
	\end{tabularx}
\end{table}

Metrics such as the number of posts, followers, friends, account creation date, and privacy setting are provided by the attributes \texttt{statuses\_count}, \texttt{followers\_count}, \texttt{friends\_count}, \texttt{created\_at}, and \texttt{protected}. Another important profile attribute is \texttt{description}. Users typically include a short description on their profile to introduce themselves or summarize their activities; analyzing this text can reveal useful information about a user's interests and persona. Twitter also exposes appearance-related attributes such as \texttt{profile\_link\_color} and \texttt{default\_profile\_image}. Examining profile colors and images can contribute additional signals for assessing account authenticity.

During the second phase, we collect a selection of each user's most recent tweets. To balance completeness with Twitter's API limits and the needs of this study, we store up to 3,250 recent tweets per account; accounts with fewer tweets naturally contribute fewer items. The API also returns metadata for each tweet; Table~\ref{tweet feature} summarizes the attributes we store alongside the tweet text.

\begin{table}[h]
	\caption{Extracted features for each tweet.\label{tweet feature}}
	{\footnotesize
		\begin{tabularx}{\textwidth}{l l l}
			\toprule
			\addlinespace[2pt]
			\multicolumn{3}{c}{Features} \\
			\addlinespace[2pt]
			\hline
			\addlinespace[2pt]
			created\_at & geo & text \\
			coordinates & truncated & place \\
			entities & contributors & source \\
			is\_quote\_status & in\_reply\_to\_status\_id & retweet\_count \\
			in\_reply\_to\_status\_id\_str & favorite\_count & in\_reply\_to\_user\_id \\
			favorited & in\_reply\_to\_user\_id\_str & retweeted \\
			in\_reply\_to\_screen\_name & possibly\_sensitive & user \\
			lang & & \\
			\addlinespace[2pt]
			\bottomrule
		\end{tabularx}
	}
\end{table}

Among the returned features, several besides the tweet text are particularly informative. The \texttt{entities} field captures hashtags, URLs, user mentions, and emojis contained in a tweet. Twitter enforces a 280-character limit per tweet; users who exceed this limit split their content across multiple tweets or threads. The \texttt{truncated} attribute indicates whether the stored text is a shortened portion of a longer tweet or part of a thread. The \texttt{possibly\_sensitive} flag denotes tweets that may contain sensitive material such as explicit language or imagery; this flag is relatively new and has been applied experimentally to some tweets, so its coverage and reliability are inconsistent.

In the third phase, we collect user reactions by gathering comments directed at each account. Because of API limits and because our interest is in opinions about accounts rather than reactions to individual posts, we do not retrieve comments for each post. Instead, we store up to the 800 most recent comments made by other users that mention or reply to a given account.
After completing the three collection phases, we label (score) each account using NewsGuard. NewsGuard assigns scores based on the criteria listed in Table~\ref{NewsGuard}. During scoring we may choose to include only a subset of criteria---such as those related to transparency---or to exclude others.\footnote{
	Our unlabeled data are publicly available at \url{https://huggingface.co/datasets/mamad97/MultiCred-Dataset}. However, due to NewsGuard's licensing and copyright restrictions, we are unable to publish the labeled data.}

\section{Our proposed method}
\label{Proposed method}

\begin{figure}[ht]
	\centering
	\captionsetup{justification=centering}
	\includegraphics[width=0.8\textwidth]{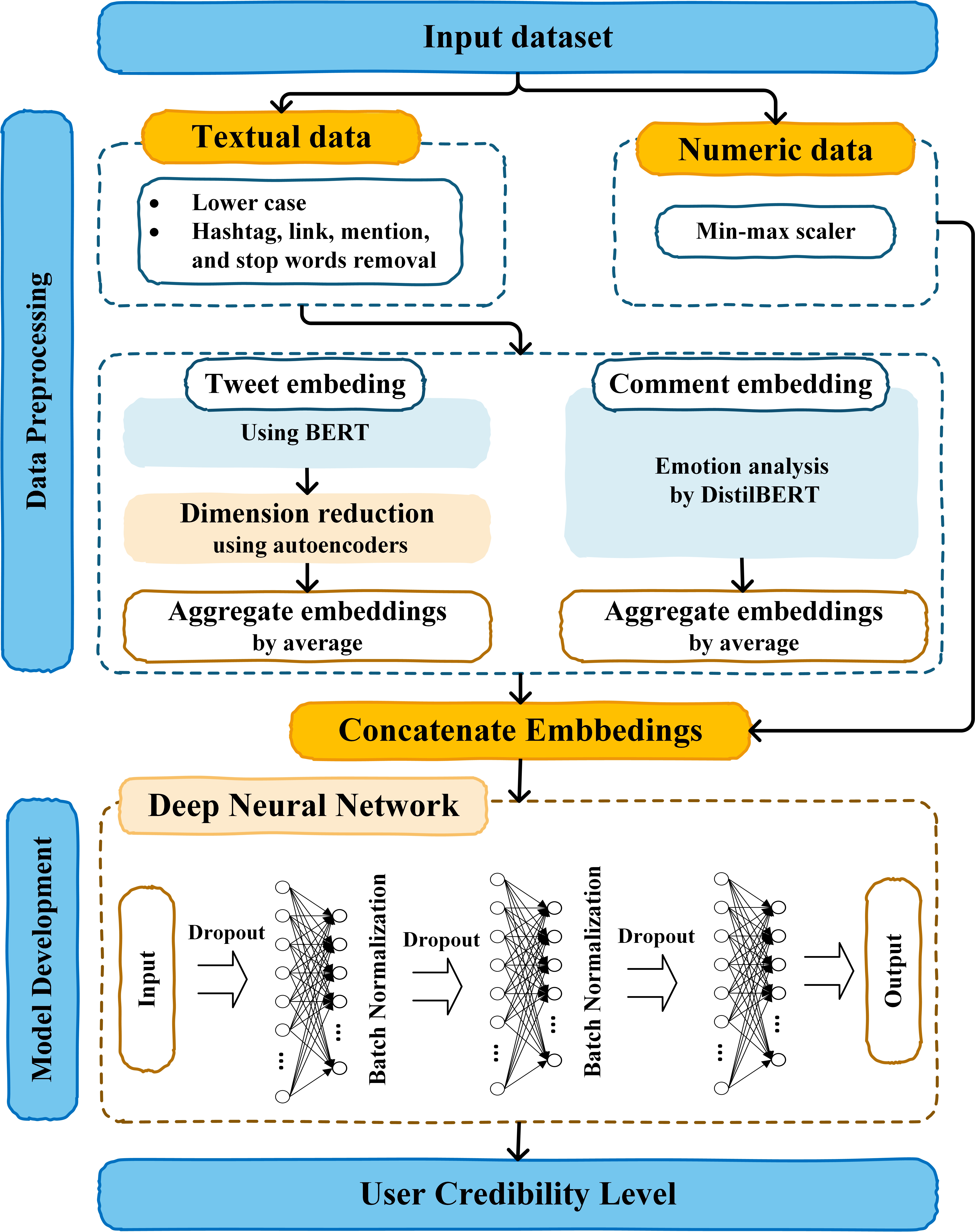}
	\caption{High-level architecture of MultiCred.\label{architecture}}
\end{figure}

In this section we present \emph{MultiCred}, a framework for assessing user credibility on social media. Figure~\ref{architecture} shows the overall pipeline and provides a high-level overview of the feature-processing and classification stages. MultiCred combines textual and non-textual features to produce a comprehensive representation of each user. The rest of this section is organized into subsections describing the framework components: preprocessing, feature extraction, feature fusion, and final classification.

\subsection{Data analysis and feature selection}	

As noted above, our dataset was collected from Twitter. The features listed in Table~\ref{Feature} are used as inputs to our model and are derived from user profiles, tweets, and comments. Preprocessing differs by feature type; the sections that follow describe the procedures applied to textual and non-textual features.

\begin{table}[h]
	\caption{Description of the features used.}
	\label{Feature}
	{\footnotesize
		\begin{tabularx}{\textwidth}{l l l l}
			\toprule
			\addlinespace[2pt]
			Feature & Type & Source & Description \\
			\midrule
			Location & Boolean & User profile & Whether the profile includes a location. \\
			Description & Boolean & User profile & Whether the profile includes a description. \\
			URL & Boolean & User profile & Whether the profile includes a URL. \\
			Protected & Boolean & User profile & Whether the profile is private. \\
			Verified & Boolean & User profile & Whether the profile is verified. \\
			Geo enabled & Boolean & User profile & Whether geolocation is enabled. \\
			Profile uses background image & Boolean & User profile & Whether a background image is used. \\
			Followers count & Numeric & User profile & Number of followers. \\
			Friends count & Numeric & User profile & Number of followings. \\
			Listed count & Numeric & User profile & Number of lists the account appears in. \\
			Statuses count & Numeric & User profile & Total number of posts. \\
			Favorite count (profile) & Numeric & User profile & Aggregate favorites (profile-level). \\
			Account creation year & Numeric & User profile & Year the account was created. \\
			Account creation month & Numeric & User profile & Month the account was created. \\
			Account creation day & Numeric & User profile & Day the account was created. \\
			Account creation hour & Numeric & User profile & Hour the account was created. \\
			Account creation minute & Numeric & User profile & Minute the account was created. \\
			Account creation second & Numeric & User profile & Second the account was created. \\
			Tweet timestamp year & Numeric & Tweet & Year of the tweet. \\
			Tweet timestamp month & Numeric & Tweet & Month of the tweet. \\
			Tweet timestamp day & Numeric & Tweet & Day of the tweet. \\
			Tweet timestamp hour & Numeric & Tweet & Hour of the tweet. \\
			Tweet timestamp minute & Numeric & Tweet & Minute of the tweet. \\
			Tweet timestamp second & Numeric & Tweet & Second of the tweet. \\
			Truncated & Boolean & Tweet & Whether the tweet is truncated or part of a thread. \\
			Retweet count & Numeric & Tweet & Number of retweets. \\
			Favorite count (tweet) & Numeric & Tweet & Number of likes for the tweet. \\
			Favorited & Boolean & Tweet & Whether the authenticated user has favorited the tweet. \\
			Retweeted & Boolean & Tweet & Whether the authenticated user has retweeted the tweet. \\
			Is quote status & Boolean & Tweet & Whether the tweet is a quote. \\
			Number of hashtags & Numeric & Tweet & Count of hashtags in the tweet. \\
			Number of user mentions & Numeric & Tweet & Count of user mentions in the tweet. \\
			Number of URLs & Numeric & Tweet & Count of URLs in the tweet. \\
			Number of symbols & Numeric & Tweet & Count of symbols (e.g., emojis) in the tweet. \\
			Poll & Boolean & Tweet & Whether the tweet contains a poll. \\
			Tweet text & Text & Tweet & Full tweet text. \\
			Comment text & Text & Comment & Full comment text. \\
			\addlinespace[2pt]
			\bottomrule
		\end{tabularx}
	}
\end{table}

\subsubsection{Non-textual features} \label{Non-textual features}

For the non-textual features in Table \ref{Feature}, we do not apply any feature selection algorithm; instead, we feed these features in their raw form directly into the prediction model. The only preprocessing step we perform is normalization. Specifically, we use min-max normalization to transform each value \(x\) into a normalized value \(\bar{x}\) as follows:
\[
\bar{x} = \frac{x - x_{\text{minimum}}}{x_{\text{maximum}} - x_{\text{minimum}}},
\]
where \(x_{\text{minimum}}\) and \(x_{\text{maximum}}\) are the minimum and maximum values, respectively, observed for that feature across all data points.

\subsubsection{Textual features}

The second category of features consists of textual data. To use these features in our learning models, we first vectorize the text, converting it into numerical vectors. Before vectorization, we apply the following preprocessing steps:
\begin{itemize}
	\item convert all text to lowercase;
	\item remove all hashtags;
	\item strip all URLs;
	\item remove all user mentions;
	\item eliminate stop words.
\end{itemize}

For vectorizing tweet texts, we utilize the BERT model \cite{DBLP:conf/naacl/DevlinCLT19}, which produces 768-dimensional representations. Such high dimensionality can complicate model training and impede convergence. To address this issue, we employ dimensionality reduction, a process that projects data from a high-dimensional space into a lower-dimensional one while retaining as much of the original information as possible. Among the available techniques, deep autoencoders have proven particularly effective for this task \cite{schmidhuber2015deep}.

Autoencoders are neural networks designed for dimensionality reduction and data compression. Each autoencoder comprises two modules: an encoder and a decoder. The encoder transforms the input data into a lower-dimensional latent representation while preserving its salient features. The decoder reconstructs the original input from this latent representation. Training is conducted in an unsupervised manner, minimizing the reconstruction error--commonly measured by Euclidean distance--between the input and the network's output.
In this work, we employ an autoencoder that accepts the 768-dimensional BERT embeddings as input and compresses them into a 10-dimensional latent space. We train this network on a dataset of 323,500 tweet embeddings. After training, we use the encoder to map new data into the learned latent space for subsequent processing.

Another category of textual features comprises comments from other users on a particular user's tweets. To leverage these opinions in the final classification, we apply sentiment analysis to the preprocessed comments. Among the available methods, we adopt the DistilBERT model \cite{DBLP:journals/corr/abs-1910-01108}, which outputs a probability distribution over emotion classes--sadness, joy, love, anger, fear, and surprise. We use these probabilities to capture other users' sentiments toward each account.

\subsubsection{Aggregation}

\begin{figure}[!ht]
	\centering
	\captionsetup{justification=centering}
	\includegraphics[width=0.7\textwidth]{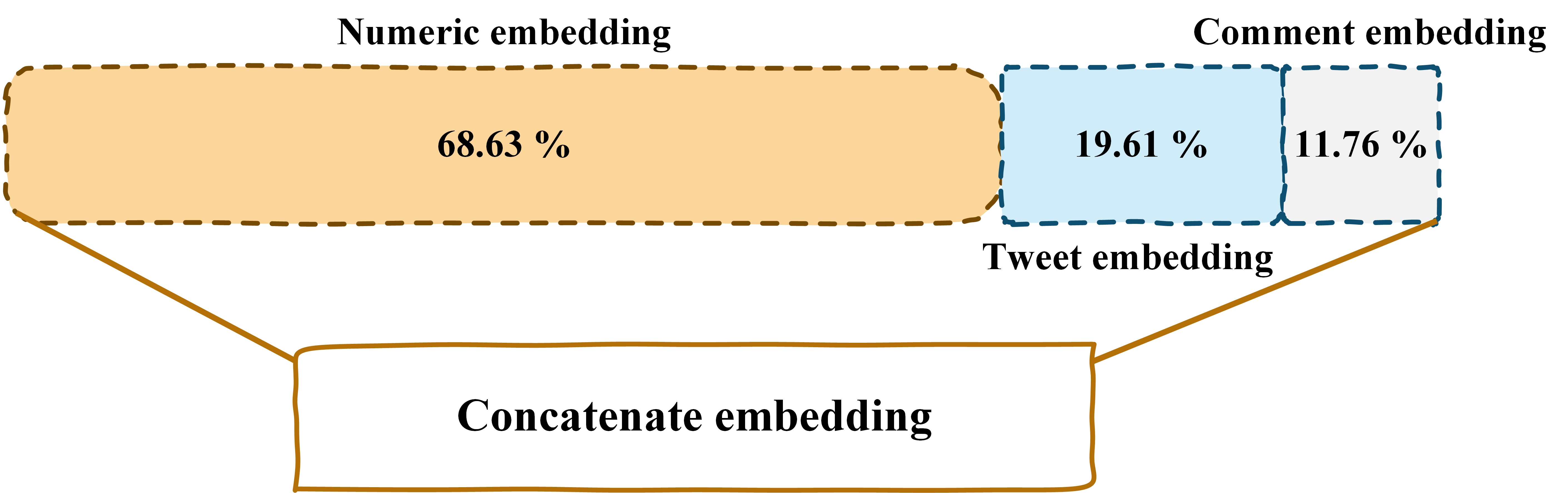}
	\caption{The final embedding created for each user.\label{embbeding}}
\end{figure}

During data collection, we retrieved 3,200 tweets and the 800 most recent comments for each user profile. The objective of preprocessing is to construct a unified vector for each account that incorporates both textual and non-textual features, enabling effective profile classification. After mapping tweets and comments into vector spaces, we aggregate each user's vectors using the mean operator. Specifically, we compute the average of a user's tweet embeddings as their tweet representation and the average of their comment embeddings as their comment representation. We then concatenate these representations with the raw non-textual feature vector to form a unique profile vector for each user. Finally, we concatenate the embedding vectors from all feature categories to generate the final feature vector for each account. Figure~\ref{embbeding} illustrates this final embedding vector and the contribution of each feature category.

\subsubsection{Labels and class imbalance} \label{Labels and class imbalance}

In the data collection phase, we assign each user a credibility score between 0 and 100, where 0 represents the lowest credibility and 100 the highest. To model this as a classification problem, we partition the continuous score range into subintervals, each corresponding to a class. In this paper, we explore multiple partitioning schemes, resulting in classification settings with varying numbers of classes.

\begin{table}	
	\caption{The number of data points for each class in
		different  classification systems.}
	\label{datapoints distribution}
	{\footnotesize
		\begin{tabularx}{\textwidth}{l l@{\hspace{17pt}} l@{\hspace{17pt}} l@{\hspace{17pt}} l@{\hspace{17pt}} l@{\hspace{17pt}} l@{\hspace{17pt}} l@{\hspace{17pt}} l@{\hspace{17pt}} l@{\hspace{17pt}} l}
			\toprule
			\addlinespace[2pt]
			Classification system & $1^{rst}$ & $2^{nd}$  & $3^{rd}$ & $4^{th}$ & $5^{th}$ & $6^{th}$ & $7^{th}$ & $8^{th}$ & $9^{th}$ & $10^{th}$ \\
			\hline
			\addlinespace[2pt]
			4-class system & $507$ & $83$ & $33$ & $24$ & - & - & - & - & - & - \\
			6-class system & $428$ & $118$ & $43$ & $29$ & $16$ & $13$ & - & - & - & - \\
			8-class system & $416$ & $91$ & $47$ & $36$ & $24$ & $9$ & $19$ & $5$ & - & - \\
			10-class system & $346$ & $137$ & $42$ & $35$ & $30$ & $16$ & $15$ & $6$ & $16$ & $4$ \\
			\bottomrule
	\end{tabularx} }
	
\end{table}

After partitioning the dataset into classification systems with varying numbers of classes, each class contains a different number of data points. Table \ref{datapoints distribution} shows the data point distribution per class across these systems, revealing that the dataset is imbalanced in every configuration. Class imbalance can bias overall performance toward majority classes and mask poor performance on minority classes. Moreover, classification accuracy often correlates with class frequency, further disadvantaging underrepresented groups. To mitigate this imbalance, we employ the Synthetic Minority Oversampling Technique (SMOTE) \cite{DBLP:journals/jair/ChawlaBHK02}.
SMOTE performs data augmentation by synthesizing new data points from existing minority-class examples. Unlike simple oversampling, it does not replicate samples exactly; instead, it creates novel instances by interpolating between a sample and one of its nearest neighbors.
Table~\ref{smote} shows the distribution of data points across different classification systems after applying SMOTE, demonstrating a substantial reduction in class imbalance.


\begin{table}
	\caption{The number of data points per class after applying the SMOTE algorithm, in different classification systems.\label{smote}} 
	{\footnotesize
		\begin{tabularx}{\textwidth}{l c}
			\toprule
			\addlinespace[2pt]
			Classification system & \hspace{2cm} Number of data points for each class \\
			\hline
			\addlinespace[2pt]
			4-class system & \hspace{2cm} $507$ \\
			6-class system & \hspace{2cm} $428$ \\
			8-class system & \hspace{2cm} $416$ \\
			10-class system & \hspace{2cm} $346$ \\
			\bottomrule
	\end{tabularx} }	
\end{table}

\subsection{The classification phase and training}

After thoroughly discussing the feature extraction and embedding learning phases, we now present our classification approach and its training procedure. Various algorithms can serve as the classification head, including traditional machine learning methods--naive Bayes, $k$-nearest neighbor, support vector machine, and random forest--as well as deep learning models such as neural networks. In this work, we employ a multi-layer neural network whose architecture is detailed in Table \ref{model architecture}.

\begin{table}[h]
	\caption{Specifications of our used neural network for classification.\label{model architecture}}
	{\footnotesize
		\begin{tabularx}{\textwidth}{l c c}
			\toprule
			\addlinespace[2pt]
			Layer(type) & Output size & \hspace{1cm} \#parameters \\
			\hline
			\addlinespace[2pt]
			dropout (Dropout) & $51$ & \hspace{1cm} $0$ \\
			hidden\_layer\_1 (Dense) & $256$ & \hspace{1cm} $13312$ \\
			batch\_normalization (BatchNormalization) & $256$ & \hspace{1cm} $1024$ \\
			dropout\_1 (Dropout) & $256$ & \hspace{1cm} $0$ \\
			hidden\_layer\_2 (Dense) & $256$ & \hspace{1cm} $65792$ \\
			batch\_normalization\_1 (BatchNormalization) & $256$ & \hspace{1cm} $1024$ \\
			dropout\_2 (Dropout) & $256$ & \hspace{1cm} $0$ \\
			hidden\_layer\_3 (Dense) & $64$ & \hspace{1cm} $16448$ \\
			Output (Dense) & $10$ & \hspace{1cm} $650$ \\
			\bottomrule
	\end{tabularx} }	
\end{table}

In this neural network, we incorporate dropout layers to prevent overfitting and batch normalization layers to accelerate the training process.
Improper weight initialization can cause training to diverge or converge slowly: excessively large initial weights may lead to exploding gradients, while excessively small weights can result in vanishing gradients. To mitigate these issues, we initialize all weights by sampling from a normal distribution.
The network contains 98,250 parameters, of which 97,226 are trainable.

During training, we split the dataset into training, testing, and validation subsets in proportions of 0.7, 0.2, and 0.1, respectively. We employ the Adam optimizer with an initial learning rate of 0.01, which decays exponentially by a factor of 0.9 over time. Data is fed into the network in mini-batches of size 16, and the rectified linear unit (ReLU) activation function is applied in every layer. Training proceeds for a maximum of 2,000 epochs, with early stopping if validation accuracy does not improve for 200 consecutive epochs.\footnote{We make the implementation of our proposed algorithm publicly available at \url{https://github.com/Mohammad-Moradi/MultiCred}.}

To provide a clear overview of the proposed \textit{MultiCred} framework, we present the entire pipeline in Algorithm~\ref{alg:multicred}. While the preceding subsections detail the processing of numeric and textual features, the following pseudocode offers a high-level description of the workflow, outlining the sequential steps from feature preprocessing to final classification. Algorithm~\ref{alg:multicred} encapsulates the core stages of MultiCred and serves as a concise reference for the overall methodology.

\begin{algorithm}[h]
	\footnotesize
	\caption{MultiCred: multilevel user credibility assessment}
	\label{alg:multicred}
	\begin{algorithmic}[1]
		\Require User numeric features $F_n$, user tweets $T = \{t_1, \dots, t_m\}$, user comments $C = \{c_1, \dots, c_k\}$
		\Ensure Credibility level $y$ for each user
		
		\State \textbf{Preprocess non-textual features:} Normalize $F_n$
		
		\State \textbf{Process tweets:}
		\For{each tweet $t_i$ in $T$}
		\State Convert text to lowercase
		\State Remove hashtags, links, usernames, stopwords
		\State $v_i \gets$ BERT representation of $t_i$
		\State $z_i \gets$ Autoencoder compress($v_i$)
		\EndFor
		\State $V_T \gets \frac{1}{m} \sum_{i=1}^{m} z_i$ \Comment{Overall tweet vector}
		
		\State \textbf{Process comments:}
		\For{each comment $c_j$ in $C$}
		\State Apply same preprocessing steps
		\State $s_j \gets$ DistilBERT sentiment vector of $c_j$
		\EndFor
		\State $V_C \gets \frac{1}{k} \sum_{j=1}^{k} s_j$ \Comment{Overall comment vector}
		
		\State \textbf{Feature fusion:} $X \gets [F_n \,\|\, V_T \,\|\, V_C]$
		\State \textbf{Classification:} $y \gets$ NeuralNetwork($X$)			
	\end{algorithmic}
\end{algorithm}

\subsection{Time complexity}
\label{Time Complexity}

To complement the description of \textit{MultiCred}, we analyze the algorithm's computational complexity. Understanding the time complexity of each stage provides insight into the efficiency and scalability of our approach, particularly given the use of BERT and autoencoder models for textual feature processing. Table~\ref{time complexity} summarizes the dominant operations and their associated time complexity for a single user, highlighting the components that contribute most to the overall computational cost.

\begin{table}[h]
	\caption{Time complexity analysis of MultiCred.\label{time complexity}}
	{\footnotesize
		\begin{tabularx}{\textwidth}{l l}
			\toprule
			\addlinespace[2pt]
			\textbf{Algorithm Step} & \textbf{Time Complexity} \\
			\midrule
			Non-textual features preprocessing (min-max normalization) & \(O(d)\) \\
			Tweets preprocessing (tokenization, cleaning) & \(O(m \cdot L_t)\) \\
			Tweets embedding (Transformer/BERT forward) & \(O(m \cdot L_t^2 \cdot H)\) \\
			Tweets compression (autoencoder inference) & \(O(m \cdot H \cdot C)\) \\
			Tweets aggregation (pooling/averaging) & \(O(m \cdot C)\) \\
			Comments preprocessing (tokenization, cleaning) & \(O(r \cdot L_c)\) \\
			Comments embedding (Transformer forward for sentiment) & \(O(r \cdot L_c^2 \cdot H_s)\) \\
			Comments aggregation (pooling/averaging) & \(O(r \cdot S)\) where \(S\) is sentiment feature size \\
			Feature fusion (concatenation of vectors) & \(O(d + C + S)\) \\
			Classification (NN forward pass) & \(O(P)\) \\
			\addlinespace[2pt]
			\textbf{Overall per user (dominant terms)} & \(O(m \cdot L_t^2 \cdot H + r \cdot L_c^2 \cdot H_s)\) \\
			\bottomrule
		\end{tabularx}
	}
\end{table}

As Table~\ref{time complexity} indicates, the primary computational bottleneck in MultiCred is the transformer-based embedding stages: generating BERT embeddings for tweets and DistilBERT embeddings for comments, both of which scale quadratically with their sequence lengths. In contrast, all other stages--preprocessing, autoencoder compression, and feature aggregation--have a relatively low impact on the overall computational cost.

\section{Empirical results} \label{Empirical results}

In this section, we evaluate the empirical effectiveness of our proposed model by comparing it with state-of-the-art credibility assessment methods. We begin by outlining the evaluation criteria, then briefly describe the benchmark methods. Finally, we present our empirical results and provide detailed discussion and analysis.

\subsection{Evaluation metrics}

To evaluate classification performance, we use precision, recall, and F1-score. Precision measures the proportion of correctly predicted positive instances among all positive predictions. Recall measures the proportion of actual positive instances that the model correctly identifies. Both metrics are computed from the counts of true positives (TP), false positives (FP), and false negatives (FN). The evaluation metrics are formally defined as follows:

\begin{equation}
	accuracy = \frac{TP + TN}{TP + FN + TN + FN}
\end{equation}
\begin{equation}
	precision = \frac{TP}{TP + FP}
\end{equation}
\begin{equation}
	recall = \frac{TP}{TP + FN}
\end{equation}
\begin{equation}
	F1-score = \frac{2*precision*recall}{precision + recall}
\end{equation}
where:
\begin{itemize}
	\item $TP$ (True Positive): data points belonging to a specific class that the model correctly predicts.
	\item $FP$ (False Positive): data points not belonging to a specific class that the model incorrectly predicts as belonging.
	\item $TN$ (True Negative): data points not belonging to a specific class that the model correctly predicts as not belonging.
	\item $FN$ (False Negative): data points belonging to a specific class that the model incorrectly predicts as not belonging.
\end{itemize}

\subsection{Baseline methods}

We adopt two recent state-of-the-art methods as baselines: Bharti and Pandey's approach \cite{DBLP:journals/soco/BhartiP21} and the method of Verma et al. \cite{DBLP:journals/snam/VermaAMG22}.

Kumari et al. \cite{DBLP:journals/soco/BhartiP21} proposed a system for identifying fake Twitter users. Their dataset comprises two components: the former was collected and manually labeled by the authors via the Twitter API, and the latter was sourced from the 2015 "TheFakeProject" study. The combined dataset includes 6,973 user accounts, of which 3,752 are labeled fake and 3,221 genuine. All features are non-textual, and multiple feature-selection techniques were applied to refine the feature set. For classification, they employed a logistic regression model with parameters optimized via particle swarm optimization.

Verma et al.~\cite{DBLP:journals/snam/VermaAMG22} proposed UCred, a framework that assesses Twitter user credibility by integrating machine learning and deep learning techniques. UCred classifies users into two categories: genuine and fake. Their dataset consists of 1,337 fake profiles and 1,481 genuine profiles. After preprocessing, models are trained in three distinct categories, the best-performing model from each category is selected, and a majority-voting mechanism among these finalists is applied to classify each user.

\subsection{Results}

In this section we present our empirical results. Two clarifications precede the findings. First, whereas most prior work---including the two baseline methods discussed earlier---frames credibility assessment as a binary task (genuine vs.\ fake), MultiCred is implemented as a multi-class classifier; we therefore adapt the baselines to the same multi-class setup. Second, because MultiCred uses SMOTE to address class imbalance, we apply SMOTE to all baseline methods as well to ensure fair comparisons.

Each experiment was repeated ten times; we report the mean and standard deviation for all metrics. Table~\ref{Result} summarizes these outcomes. As shown in Table~\ref{Result}, MultiCred significantly outperforms all competing methods across every classification setting and evaluation metric.

\begin{table}[h]
	\caption{Comparing the performance of MultiCred against the baseline algorithms in different classification systems.}
	\label{Result}
	{
		\footnotesize
		\begin{tabularx}{\textwidth}{l l l l l l}
			\toprule
			\addlinespace[2pt]
			Classification system & Model & Precision(\%) & Recall(\%) & F1-score(\%) & Accuracy(\%) \\
			\addlinespace[2pt]
			\hline
			\addlinespace[2pt]
			\multirow{3}{*}{4-class} & Method of \cite{DBLP:journals/soco/BhartiP21} & $41.92\pm1.76$ & $42.13\pm1.70$ & $41.79\pm1.61$ & $40.01\pm1.68$ \\
			\addlinespace[2pt]
			& Method of \cite{DBLP:journals/snam/VermaAMG22} & $75.18\pm3.32$ & $72.77\pm1.95$ & $73.82\pm2.62$ & $72.77\pm1.90$ \\
			\addlinespace[2pt]
			
			& MultiCred & \textbf{89.56$\pm$1.69}  & \textbf{89.09$\pm$1.61} & \textbf{88.89$\pm$1.60} & \textbf{89.35$\pm$1.52} \\
			
			\cmidrule(lr){2-6}
			
			\multirow{3}{*}{6-class} & Method of \cite{DBLP:journals/soco/BhartiP21} & $29.43\pm2.01$ & $30.02\pm1.91$ & $28.93\pm2.37$ & $30.1\pm1.91$ \\
			\addlinespace[2pt]
			& Method of \cite{DBLP:journals/snam/VermaAMG22} & $31.74\pm4.19$ & $39.46\pm3.81$ & $34.31\pm4.12$ & $39.46\pm3.85$ \\
			\addlinespace[2pt]
			& MultiCred & \textbf{85.10$\pm$4.23} & \textbf{85.63$\pm$3.78} & \textbf{84.88$\pm$4.25} & \textbf{86.14$\pm$3.71} \\
			
			\cmidrule(lr){2-6}
			
			\multirow{3}{*}{8-class} & Method of \cite{DBLP:journals/soco/BhartiP21} & 27.27$\pm$1.89 & 29.60$\pm$1.42 & 27.45$\pm$1.37 & 29.34$\pm$1.30 \\
			\addlinespace[2pt]
			& Method of \cite{DBLP:journals/snam/VermaAMG22} & 29.78$\pm$6.5 & 35.61$\pm$4.99 & 31.80$\pm$6.19 & 34.69$\pm$4.88 \\
			\addlinespace[2pt]
			
			& MultiCred & \textbf{86.60$\pm$2.86} & \textbf{86.78$\pm$2.31} & \textbf{85.92$\pm$2.29} & \textbf{86.71$\pm$2.36} \\
			
			\cmidrule(lr){2-6}
			
			\multirow{3}{*}{10-class} & Method of \cite{DBLP:journals/soco/BhartiP21} & 26.05$\pm$2.43 & 28.09$\pm$1.54 & 25.78$\pm$1.90 & 28.14$\pm$1.60 \\
			\addlinespace[2pt]
			& Method of \cite{DBLP:journals/snam/VermaAMG22} & 46.51$\pm$2.31 & 45.30$\pm$1.43 & 45.78$\pm$1.66 & 45.30$\pm$1.36 \\
			\addlinespace[2pt]
			& MultiCred & \textbf{86.89$\pm$1.30} & \textbf{87.29$\pm$0.97} & \textbf{85.85$\pm$1.08} & \textbf{87.61$\pm$0.83} \\
			
			\bottomrule
	\end{tabularx} }
\end{table}

\subsection{Analysis}

We assess individual feature contributions using Shapley Additive Explanations (SHAP) \cite{DBLP:conf/nips/LundbergL17}. SHAP leverages Shapley values from cooperative game theory to assign each feature an importance value for a given prediction. Unlike many traditional importance measures, SHAP yields consistent, theoretically grounded attributions and supports both local (instance-level) and global (dataset-level) interpretability. We use SHAP to produce transparent explanations that quantify how each feature influences the model's outputs.

First, we quantify the overall importance of three feature categories---profile features, tweet embeddings, and comment sentiment embeddings---by summing the absolute SHAP values across all samples and model classes. Figure~\ref{overall_feature_importance} presents these results as bar plots, highlighting each category's relative contribution to the model's predictions. This high-level summary clarifies which information source---profile data, tweet content, or comment sentiment---exerts the greatest influence on the classifier.

\begin{figure}[!ht]
	\centering
	\captionsetup{justification=centering}
	\includegraphics[width=\textwidth]{overall.png}
	\caption{Overall importance of feature categories measured by summed absolute SHAP values.\label{overall_feature_importance}}
\end{figure}

We next examine profile features in detail. For each credibility class, we compute absolute SHAP values for individual profile features and aggregate them by class. Figure~\ref{profile_feature_importance} displays these results as bar plots, showing each feature's relative contribution to predicting its corresponding class. This class-level analysis identifies which profile attributes, such as engagement and activity metrics, most strongly distinguish between credibility levels.

\begin{figure}[!ht]
	\centering
	\captionsetup{justification=centering}
	\includegraphics[width=\textwidth]{profile_importance.png}
	\caption{Importance of individual profile features per credibility class measured by summed absolute SHAP values.\label{profile_feature_importance}}
\end{figure}

The feature-importance analysis yields several insights. At the category level, \emph{profile features} provide the strongest signal for credibility, followed by \emph{tweet embeddings} and \emph{comment sentiment embeddings}, indicating user metadata is most predictive. The class-specific analysis shows \emph{account creation time} is the single most influential feature across classes. Other features with large overall importance---\emph{number of symbols}, \emph{presence of profile URL}, and \emph{geo-enabled}---contribute primarily to a single class. \emph{Engagement metrics} (likes, favorites count, favorited status, retweet count) are consistently important for distinguishing credibility levels. By contrast, traditional indicators used in binary fake-account detection---\emph{statuses count}, \emph{verified status}, and \emph{follower count}---have a more balanced but smaller impact in this multi-class setting. Overall, both global and class-specific features shape model decisions, with engagement and account metadata being the strongest predictors.

Considering a broad spectrum of features and processing them appropriately improves MultiCred's performance relative to using only a subset. Experiments show that combining textual and non-textual features yields better results: adding user opinions (comment sentiment embeddings) to the final feature vector increases accuracy by an average of 4.09\% across all classification settings, and including tweet embeddings produces an additional average accuracy gain of 2.46\%.

\begin{figure}[!ht]
	\centering
	\captionsetup{justification=centering}
	\includegraphics[width=\textwidth]{comparison_3.png}
	\caption{Accuracy comparison of MultiCred and two baseline methods across varying numbers of classes.\label{comparison}}
\end{figure}

F1 scores decline for all methods as the number of classes increases because class boundaries become more entangled in feature space, making correct discrimination harder. MultiCred experiences the same downward trend but consistently outperforms the baselines and achieves a substantial relative improvement across class counts.

Figure~\ref{comparison} shows that the method of Bharti and Puri \cite{DBLP:journals/soco/BhartiP21} experiences only a slight decline in evaluation metrics when moving from eight to ten classes. UCred \cite{DBLP:journals/snam/VermaAMG22} actually improves on several metrics in the ten-class setting. MultiCred likewise records modest gains in recall, precision, and accuracy for ten classes versus eight, suggesting that all three methods can retain or even enhance performance under a finer-grained class partitioning.

This pattern indicates the dataset is better represented by a ten-class scheme: the feature-space distribution supports finer-grained distinctions, and users align more naturally with ten credibility levels than eight. Adopting additional credibility tiers therefore yields a more nuanced and often more accurate assessment of user credibility.

\section{Conclusion and future work}
\label{Conclusion and future work}

In this paper, we studied the problem of multilevel user credibility assessment in social networks. We first collected a dataset tailored for evaluating credibility across multiple levels. We then proposed the MultiCred model, which assigns users to one of several credibility tiers based on a rich and diverse set of features extracted from their profiles, tweets, and comments. MultiCred leverages deep language models for textual analysis and deep neural networks for non-textual data processing. Our experiments on the collected dataset demonstrate that MultiCred outperforms existing methods across various accuracy metrics.

Due to computational constraints, this study did not include certain feature types--such as images and other multimedia content shared by users. We also excluded the social network graph, which encodes valuable structural information about user interactions. Integrating this graph data with our existing features could yield deeper insights into user behavior. Exploring these additional modalities and incorporating them into the MultiCred framework is a promising direction for future work.

%
%
%
%
%
%
%
%
%


\bibliographystyle{plain}
\bibliography{site}

\end{document}